\documentclass{emulateapj}

\begin{document}

\shortauthors{Luhman}
\shorttitle{250 K Brown Dwarf at 2 pc}

\title{Discovery of a $\sim250$~K Brown Dwarf at 2~pc from the Sun\altaffilmark{1}}

\author{K. L. Luhman\altaffilmark{2,3}}

\altaffiltext{1}
{Based on data from the {\it Wide-field Infrared Survey Explorer},
the {\it Spitzer Space Telescope}, Gemini Observatory, and
the VISTA Telescope at ESO's Paranal Observatory.}

\altaffiltext{2}{Department of Astronomy and Astrophysics,
The Pennsylvania State University, University Park, PA 16802, USA;
kluhman@astro.psu.edu}
\altaffiltext{3}{Center for Exoplanets and Habitable Worlds, The 
Pennsylvania State University, University Park, PA 16802, USA}

\begin{abstract}

Through a previous analysis of multi-epoch astrometry from the
{\it Wide-field Infrared Survey Explorer (WISE)}, I identified
WISE J085510.83$-$071442.5 as a new high proper motion object.
By combining astrometry from {\it WISE} and the {\it Spitzer Space Telescope},
I have measured a proper motion of $8.1\pm0.1\arcsec$~yr$^{-1}$ and a
parallax of $0.454\pm0.045\arcsec$ ($2.20^{+0.24}_{-0.20}$~pc) for
WISE J085510.83$-$071442.5, giving it the third highest proper motion 
and the fourth largest parallax of any known star or brown dwarf.
It is also the coldest known brown dwarf based on its absolute magnitude at
4.5~\micron\ and its color in $[3.6]-[4.5]$.
By comparing $M_{4.5}$ with the values predicted by theoretical
evolutionary models, I estimate an effective temperature of 225--260~K
and a mass of 3--10~$M_{\rm Jup}$ for the age range of 1--10~Gyr that
encompasses most nearby stars.

\end{abstract}


\section{Introduction}

The closest stars to the Sun have played a central role in studies of stellar
astrophysics, as well as appealing to the imagination of the general
public \citep{hen97}. Over the last century, wide-field imaging surveys have 
been conducted at progressively fainter magnitudes and longer wavelengths,
enabling the detection of the Sun's neighbors down to the hydrogen burning
limit \citep[e.g.,][]{bar16,wol19,ros26,luy79,lep05b} and into the
substellar regime \citep[e.g.,][]{kir99,str99,bur04,burn10}.
One of the most recent surveys was performed by the {\it Wide-field
Infrared Survey Explorer} \citep[{\it WISE},][]{wri10}, which obtained
mid-infrared (IR) images of the entire sky. Those data have proven to be
highly effective at uncovering the coldest brown dwarfs in the vicinity of
the Sun \citep{cus11,kir11}.

Most of the brown dwarfs found with {\it WISE} have been selected based
on their colors. However, nearby brown dwarfs also can be identified through
their large proper motions, which avoids photometric selection biases
\citep[e.g.,][]{dea09,she09,art10,kir10,sch10}.
This method has been successfully applied to the {\it WISE} astrometry
by measuring motions relative to near-IR surveys
\citep{giz11,liu11,sch11,bih13} and within the
multiple epochs from {\it WISE} \citep{luh13,luh14,tho13,wri14,kir14}.
The latter data are especially well-suited for finding nearby objects that
are too cold to be detected in near-IR surveys.
In this Letter, I characterize an object of this kind from my proper motion
survey in \citet{luh14}\footnote{In a subsequent study, \citet{kir14} also
independently identified this high proper motion object.},
demonstrating that it is one of the Sun's closest neighbors and the coldest
known brown dwarf.

\section{Observations}

\subsection{Mid-IR Images from {\it WISE}}
\label{sec:wise}

Between 2010 January 7 and 2011 February 1, 
{\it WISE} performed an all-sky imaging survey
in bands centered at 3.4, 4.6, 12, and 22~\micron,
which are denoted as $W1$, $W2$, $W3$, and $W4$, respectively \citep{wri10}.
Because of the successive depletion of the two cryogen tanks,
images were collected through only W1/W2/W3 and W1/W2 after 2010 August 6
and 2010 September 29, respectively.
Each position in the sky was observed $\gtrsim12$ times
over a period of $\gtrsim1$~day at intervals of six months.
By the end of the 13 month survey, the images for a given location spanned
either 6 or 12 months.

\begin{figure}
\epsscale{1}
\plotone{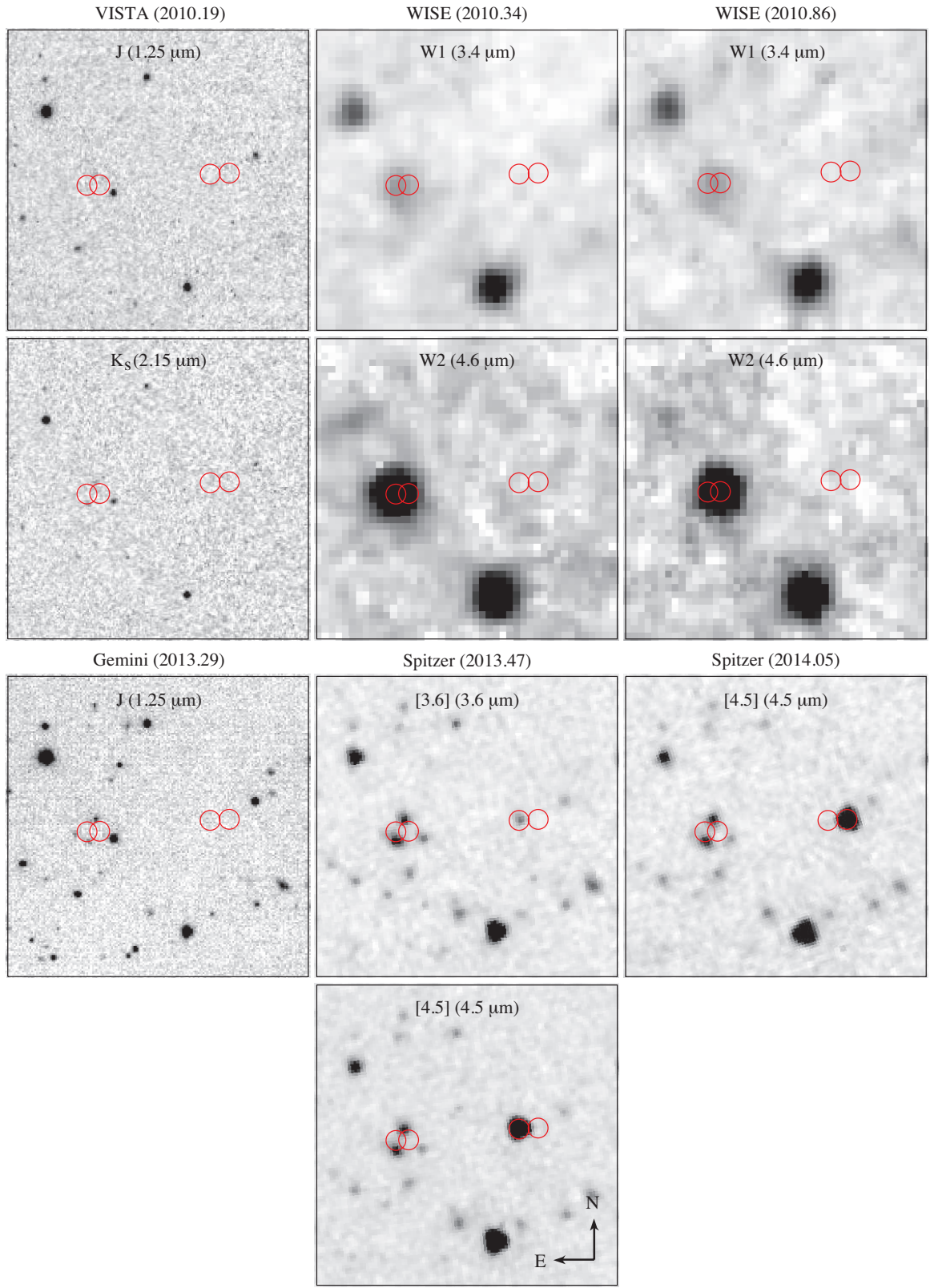}
\caption{
Images of WISE~0855$-$0714 from VISTA, {\it WISE}, Gemini, and {\it Spitzer}.
In the {\it WISE} images, WISE~0855$-$0714 is a blend of a moving object
that dominates at $W2$ and two stationary sources that likely dominate at $W1$.
The circles indicate the positions of the moving component 
in the {\it WISE} and {\it Spitzer} images; it is not detected by VISTA
or Gemini.  The size of each image is $1\arcmin\times1\arcmin$.
}
\label{fig:image}
\end{figure}

In \citet{luh14}, I found that WISE J085510.83$-$071442.5
(hereafter WISE~0855$-$0714) moved $2\farcs5$ between
two epochs of {\it WISE} images that are separated by six months,
which indicates a proper motion that is unusually high among known stars.
The {\it WISE} data for this object were obtained on 2010 May 4
and 2010 November 11 and consisted of 14 and 13 detections, respectively.
It has a color of $W1-W2=2.7$ in the {\it WISE} All-Sky Source Catalog,
which implies a spectral type of T6--T8 \citep{kir11}.
Among the publicly available images that encompass WISE~0855$-$0714,
the data at $J$ and $K_s$ from the Visible and Infrared Survey Telescope for
Astronomy (VISTA) Hemisphere Survey (PI. McMahon, ID 179.A-2010) provide
the best constraints on its nature. These images are found in
the first public data release for that survey in the
VISTA Science Archive. The VISTA observations were performed
on the night of 2010 March 9, which was only two months prior to the
first epoch of {\it WISE} images for WISE~0855$-$0714.
As a result, its expected position in the VISTA images is well-constrained.
An object is not detected at that location in the $J$ and $K_s$
data from VISTA, which are shown in Figure~\ref{fig:image}
for a $1\arcmin\times1\arcmin$ area that encompasses WISE~0855$-$0714.
Using the VISTA catalog of sources in the vicinity
of WISE~0855$-$0714, I estimate that a signal-to-noise ratio (SNR) of 3
corresponds to $J\sim20.3$ and $K_s\sim18.6$. The resulting color of $J-W2>7$
is much redder than the values of 2.5--4 that are exhibited by
T6--T8 dwarfs, and instead is indicative of a Y dwarf \citep{kir11}.

To investigate the discrepancies in the spectral types implied by $W1-W2$
and $J-W2$, I examined the {\it WISE} images of WISE~0855$-$0714.
I retrieved the single exposure images from the NASA/IPAC Infrared Science
Archive, and coadded all images at a given filter for each of the two epochs.
The resulting images at $W1$ and $W2$ are shown in Figure~\ref{fig:image}.
The mean coordinates of WISE~0855$-$0714 from the single exposure
catalogs in each epoch agree with the positions of the $W2$ source in the
coadded images, but each epoch's source in $W1$ is near the midpoint between
those coordinates. In other words, the $W2$ source shows significant movement
between epochs while the $W1$ source does not.
The position of WISE~0855$-$0714 in $W1$ is roughly midway between
two objects that are near the detection limit in the VISTA $K_s$ image
and are separated by 4--$5\arcsec$. Based on these {\it WISE} and VISTA data,
I concluded that WISE~0855$-$0714 is probably a blend of a moving object
that dominates at $W2$ and two stationary sources that dominate at $W1$,
which would explain why $W1-W2$ implies an earlier spectral type than $J-W2$.
This scenario is confirmed by the {\it Spitzer} observations described
in Section~\ref{sec:irac}.
It is unclear which components of WISE~0855$-$0714
dominate in $W3$, which is available for the first epoch only.
The {\it WISE} All-Sky Source Catalog contains a measurement of photometry
for WISE~0855$-$0714 in $W4$, but no detection is apparent based on visual
inspection of the coadded image. 

\subsection{Near-IR Images from Gemini}

I pursued near-IR imaging of WISE~0855$-$0714 that is deeper than the data
from VISTA to better constrain its motion and spectral type and to assess
the feasibility of spectroscopy.
Among the standard near-IR filters, $J$ and $H$ usually offer the best
sensitivity to the coldest brown dwarfs; the former was selected for
these observations. The images were obtained with the Gemini Near-Infrared
Imager (NIRI) at the Gemini North telescope on the night of 2013 April 17.
The instrument contained a 1024$\times$1024
ALADDIN InSb array and was operated with the f/6 camera, resulting
in a plate scale of $0\farcs117$~pixel$^{-1}$ and a field of view of
$2\arcmin\times2\arcmin$. Twenty-six dithered images of WISE~0855$-$0714
were collected, each with an exposure time of 1~min.
The images of WISE~0855$-$0714 were flat fielded, corrected for distortion,
registered, and combined. Astrometry and photometry from VISTA were used
to measure the world coordinate system (WCS) and flux calibration for the
combined image. Point sources in the image exhibit FWHM$\sim0\farcs5$.
No counterpart to the moving component of WISE~0855$-$0714
is detected in the NIRI image (see Fig.~\ref{fig:image}).
The two components detected by VISTA
appear at the same positions in the NIRI image, confirming that they
have negligible motion. Using the calibrated photometry for sources in
the NIRI image, I estimate that SNR=3 corresponds to $J\sim23$.

\subsection{Mid-IR Images from {\it Spitzer}}
\label{sec:irac}

Because it was likely to be a new member of the solar neighborhood based
on its rapid motion in the {\it WISE} images, I included WISE~0855$-$0714
in a survey for common proper motion companions to nearby stars
(program 90095) with the {\it Spitzer Space Telescope} \citep{wer04}.
For WISE~0855$-$0714, the data from this program also provide
constraints on its colors and motion.
It was observed on 2013 June 21 and 2014 January 20 with {\it Spitzer}'s
Infrared Array Camera \citep[IRAC;][]{faz04}. The images in June were collected
through the 3.6 and 4.5~\micron\ filters, which are denoted as [3.6] and [4.5].
Because the primary purpose of the second observation was astrometry,
only the [4.5] band was employed.
Five dithered images were obtained in each filter on a given date.
The exposure times for the individual frames were 23.6 and
26.8~s for [3.6] and [4.5], respectively. 

The IRAC data were reduced in the manner described by \citet{luh12}.
The resulting images are shown in Figure~\ref{fig:image}.
IRAC has detected two objects straddling the position
of WISE~0855$-$0714, which coincide with the pair of stationary sources
detected by VISTA and NIRI. In the first and second IRAC epochs,
the moving component of WISE~0855$-$0714 appears $\sim22\arcsec$ and
$26\arcsec$ west of its position in the second epoch of {\it WISE} images,
respectively.
This object is fainter than the two stationary sources in [3.6], but the
opposite is true at [4.5]. 
Given the similarities in the bandpasses of [3.6]/$W1$ and [4.5]/$W2$,
these relative IRAC fluxes confirm the suggestion from Section~\ref{sec:wise}
that WISE~0855$-$0714 is dominated by a moving object in $W2$ and
two stationary objects in $W1$. In fact, because $W1$ is fainter than
[3.6] for T and Y dwarfs \citep{kir11}, the dominance of the stationary
objects at $W1$ should be even greater than that in [3.6].

I measured photometry in [3.6] and [4.5] for the moving and stationary
components of WISE~0855$-$0714 from the IRAC images using the methods
from \citet{luh12}. Those data are presented in Table~\ref{tab:data}.
The $W2$ and [4.5] filters produce photometric magnitudes that agree to within
a few percent on
average\footnote{http://wise2.ipac.caltech.edu/docs/release/allsky/expsup/sec6\_3c.html}.
Therefore, I can estimate the $W2$ photometry of the moving component of
WISE~0855$-$0714 by subtracting the [4.5] photometry of the stationary objects
from the $W2$ measurement in the All-Sky Source Catalog ($W2=13.63$).
The corrected photometry for the moving component is $W2=13.89$,
which matches the IRAC measurement of $[4.5]=13.89$.

To measure astrometry for the moving component of WISE~0855$-$0714,
I began by measuring pixel coordinates with the task
{\it starfind} in IRAF for all point sources in the reduced [4.5] image
from each IRAC epoch and in the coadded $W2$ image from each {\it WISE} epoch
for a $8\arcmin\times8\arcmin$ field centered on WISE~0855$-$0714.
I used astrometry from the Two Micron All-Sky Survey \citep[2MASS,][]{skr06} 
for sources in each image to derive offsets in right ascension, declination,
and rotation that align the WCS with the 2MASS astrometric system.
For each of the {\it WISE} images, the coordinates measured for
WISE~0855$-$0714 apply to a blend of the moving object and the two stationary
sources. To correct for the latter, I added an artificial star with the flux 
of the moving component to one of the IRAC [4.5] images at a location near
the coordinates measured from a given {\it WISE} epoch, smoothed the
resulting image to the resolution of {\it WISE}, and measured astrometry
for the blended source. This process was repeated for artificial stars at
different locations 
until the astrometry of the blend agreed with the {\it WISE} position.
I then adopted the true coordinates of the artificial star for that best fit,
which was equivalent to adding $0\farcs2$ and $-0\farcs3$ to the right
ascensions measured from the first and second epochs of {\it WISE}
images, respectively. No corrections to the declinations were necessary.
To estimate the errors in the astrometry, I computed the standard
deviations of the differences in right ascension and declination
between the two IRAC epochs and between each {\it WISE} epoch and 2MASS
for sources that have fluxes roughly similar to that of WISE~0855$-$0714.
The implied errors for the {\it WISE} data are $\sim0\farcs2$, but I have
adopted $0\farcs4$ in an attempt to account for the additional
errors introduced by the blending.
The astrometric data for the moving component of WISE~0855$-$0714 from
{\it WISE} and IRAC are listed in Table~\ref{tab:astro}.
The coordinates of the stationary components in the IRAC images are included
as well.
The designation ``WISE~0855$-$0714" refers to the moving object alone in the
remainder of this study.

\section{Characterization of WISE 0855$-$0714}

\subsection{Parallax and Proper Motion}
\label{sec:astro}

\begin{figure}
\epsscale{1.2}
\plotone{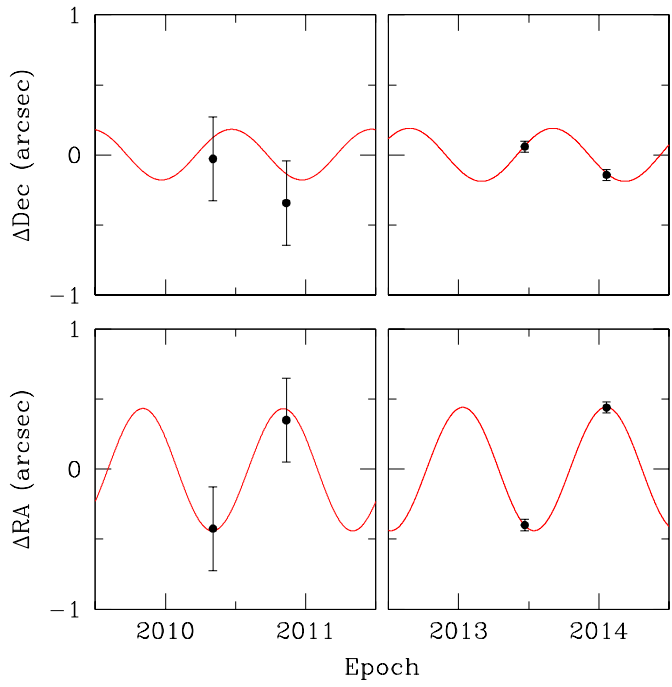}
\caption{
Relative astrometry of WISE~0855$-$0714 in images from {\it WISE} (2010) and
{\it Spitzer} (2013--2014) compared to the best-fit model of parallactic
motion (Table~\ref{tab:data}, red curve).
The proper motion produced by the fitting has been subtracted.
}
\label{fig:pm}
\end{figure}

I applied least-squares fitting of proper and parallactic motion to the
astrometry for WISE~0855$-$0714 with the IDL program MPFIT.
The reduced $\chi^2$ for the fit is less than unity (0.3), indicating
that the adopted astrometric errors may be overestimated.
To check the errors produced by the fitting, I created 1000 sets of
astrometry by adding Gaussian noise to the measured values, and fitted
parallactic and proper motion to each set.
The resulting standard deviations of proper motion and parallax
were similar to the errors from MPFIT. The estimates of proper motion and
parallax are presented in Table~\ref{tab:data}.
The relative coordinates among the four epochs are shown
in Figure~\ref{fig:pm} after subtraction of the best-fit proper motion.

\subsection{Photometric and Physical Properties}
\label{sec:phot}

\begin{figure}
\epsscale{1}
\plotone{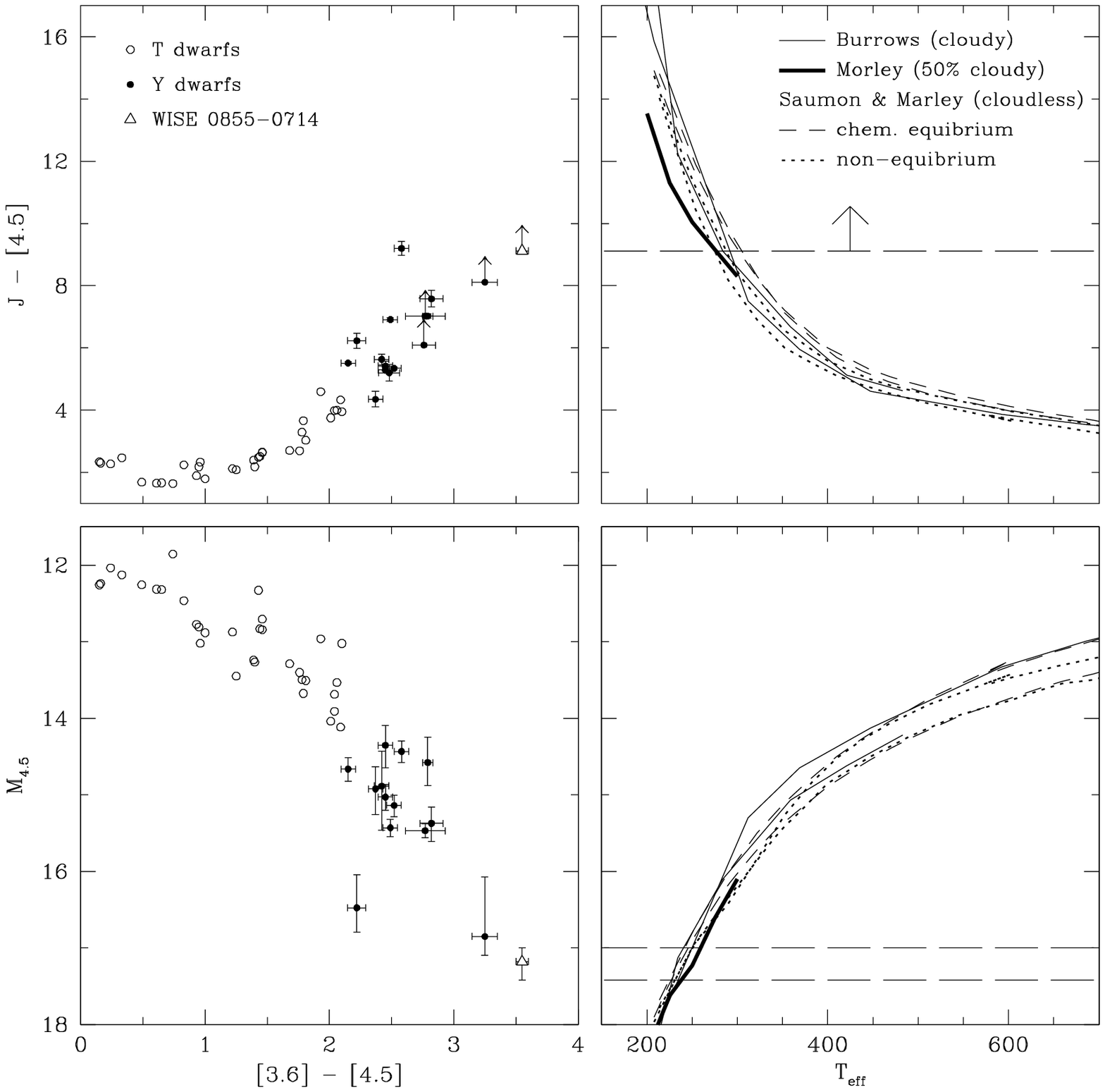}
\caption{
{\it Left}:
Color-magnitude and color-color diagrams for WISE~0855$-$0714 (open triangle)
and samples of T dwarfs \citep[open circles,][references therein]{dup12}
and Y dwarfs \citep[filled circles,][]{cus11,luh12,tin12,bei13,bei14,leg13,mar13,kir13,dup13}.
Error bars are included for WISE~0855$-$0714 and the Y dwarfs. {\it Right}:
Predicted $J-[4.5]$ and $M_{4.5}$ as a function of effective temperature
from \citet[][solid lines, 1 and 5~Gyr]{bur03},
\citet[][short dashed and dotted lines, 1 and 10 Gyr]{sau08}, 
and \cite[][thick solid lines, log~$g=4.0$, $\sim1$--3~Gyr]{mor14}.
The constraints on $J-[4.5]$ and $M_{4.5}$ for WISE~0855$-$0714 are
indicated (long dashed lines).
}
\label{fig:cmd}
\end{figure}

To characterize the photometric properties of WISE~0855$-$0714,
I have placed it in diagrams of $J-[4.5]$ versus $[3.6]-[4.5]$ and
$M_{4.5}$ versus $[3.6]-[4.5]$ in Figure~\ref{fig:cmd}.
For comparison, I have included data for known T and Y dwarfs with
measured parallaxes and photometry in these bands
\citep{cus11,dup12,luh12,tin12,bei13,bei14,leg13,mar13,kir13,dup13}.
In Figure~\ref{fig:cmd}, WISE~0855$-$0714 is the reddest brown
dwarf in $[3.6]-[4.5]$ and a contender for the reddest in $J-[4.5]$
with a few Y dwarfs that also lack detections in $J$.
None of the confirmed brown dwarfs that are absent from Figure~\ref{fig:cmd}
(e.g., no parallax measurements) have measured $[3.6]-[4.5]$ 
or $J-[4.5]$ that are redder than the colors of WISE~0855$-$0714.
It is also the faintest known brown dwarf in $M_{4.5}$.
These photometric properties indicate that WISE~0855$-$0714
is the coldest known brown dwarf.

I have estimated the effective temperature of WISE~0855$-$0714 by
comparing its $J-[4.5]$ and $M_{4.5}$ to the predictions
of theoretical evolutionary models of brown dwarfs. The color
$[3.6]-[4.5]$ was omitted from this exercise because significant
errors are likely present in the theoretical fluxes at [3.6] \citep{leg10a}.
I have used the cloudy models from \citet{bur03}
and the versions of the cloudless models from \citet{sau08} 
that were utilized by \citet{luh12} in a similar analysis of WD~0806-661~B.
I have also employed models by \citet{mor14} that include clouds of sulfides,
alkali salts, and water ice across 50\% of the surface.
Since the age of WISE~0855$-$0714 is unknown, I wish to compare it to model
predictions for a span of ages that encompasses most stars in the
solar neighborhood, such as 1--10~Gyr. Therefore, I consider the models for
1 and 10~Gyr from \citet{sau08} and the models for 1 and 5~Gyr from
\citet{bur03} (who did not perform calculations beyond 5~Gyr).
For $T<300$~K, the curve of $M_{4.5}$ versus temperature from \citet{mor14}
changes very little with age, so I use their models for a single surface
gravity of log~$g=4.0$ ($\sim1$--3~Gyr for 200--300~K).
The predicted values of $J-[4.5]$ and $M_{4.5}$
are plotted as a function of temperature in Figure~\ref{fig:cmd}.
The constraints on $J-[4.5]$ and $M_{4.5}$ for WISE~0855$-$0714
indicate temperatures of $\lesssim300$~K and 225--260~K, respectively,
based on the full set of models, and $\lesssim275$~K and 240--260~K 
according to the new calculations from \citet{mor14}.

The mass of WISE~0855$-$0714 can be estimated by comparing the observed
and theoretical values of $M_{4.5}$.
The combination of the models from \citet{bur03} and \citet{sau08}
enable an estimate of the full range of possible masses for 1--10~Gyr.
The faint limit for $M_{4.5}$ corresponds to 3~$M_{\rm Jup}$ using
the models from \citet{bur03} for 1~Gyr while the bright limit implies
10~$M_{\rm Jup}$ according to the models from \citet{sau08} for 10~Gyr. 

\section{Discussion}

WISE~0855$-$0714 has the third highest proper motion of any known object
outside the solar system ($\mu=8\farcs1$~yr$^{-1}$), behind
only Barnard's star \citep[][$\mu=10\farcs3$~yr$^{-1}$]{bar16} and
Kapteyn's star \citep[][$\mu=8\farcs6$~yr$^{-1}$]{kap97}.
The four closest systems to the Sun known prior to this study are
$\alpha$ Cen AB and Proxima Cen \citep[1.338$\pm$0.002,
1.296$\pm$0.004~pc,][]{sod99,van07},
Barnard's star \citep[1.834$\pm$0.001~pc,][]{ben99},
WISE J104915.57$-$531906.1~AB \citep[2.02$\pm$0.02~pc,][]{luh13,bof14}, and
Wolf 359 \citep[2.386$\pm$0.012~pc,][]{van95}.
With a parallactic distance of $2.20^{+0.24}_{-0.20}$~pc,
WISE~0855$-$0714 likely ranks fourth in proximity to the Sun.

Among known T and Y dwarfs, WISE~0855$-$0714 is the reddest in $[3.6]-[4.5]$,
a contender for the reddest in $J-[4.5]$, and the faintest in $M_{4.5}$,
indicating that it is the coldest known brown dwarf (and hence a Y dwarf).
When compared to the model predictions of \citet{bur03},
\citet{sau08}, and \citet{mor14}, the constraints on 
$J-[4.5]$ and $M_{4.5}$ imply effective temperatures of
$\lesssim300$~K and 225--260~K, respectively.
If it is within the age range of 1--10~Gyr that encompasses most nearby stars,
then it should have a mass
of 3--10~$M_{\rm Jup}$ according to the theoretical values of $M_{4.5}$.
At this mass, WISE~0855$-$0714 could be either a brown dwarf or a gas giant
planet that was ejected from its system.
The former seems more likely given that the frequency of planetary-mass
brown dwarfs is non-negligible\footnote{This is based on surveys of
star-forming regions \citep[e.g.,][]{alv12}, which are young enough that
planet ejection is unlikely to have occurred at a significant level.}
while the frequency of ejected planets is unknown.
Assuming that WISE~0855$-$0714 is a Y dwarf, the 
four closest known systems now consist of two M dwarfs and one member of every
other spectral type from G through Y.

WISE~0855$-$0714 offers an opportunity to test atmospheric models in an
unexplored temperature regime. 
Exploiting this opportunity will require additional astrometry to refine its
parallax measurement and deeper near-IR photometry to
better constrain its spectral energy distribution.
Spectroscopy will be necessary for detailed tests of the model atmospheres,
but given the current limits on the near-IR fluxes ($J>23$), it may not be
feasible until the deployment of the {\it James Webb Space Telescope}.

\acknowledgements
I acknowledge support from grant NNX12AI47G from the NASA Astrophysics
Data Analysis Program. I thank Caroline Morley and Didier Saumon for
providing their model calculations.
{\it WISE} is a joint project of the University of California, Los Angeles,
and the Jet Propulsion Laboratory (JPL)/California Institute of
Technology (Caltech), funded by NASA.
The Gemini data were obtained through program GN-2013A-DD-3.
Gemini Observatory is operated by the
Association of Universities for Research in Astronomy, Inc., under a
cooperative agreement with the NSF on behalf of the Gemini partnership:
the National Science Foundation (United States), the National Research Council
(Canada), CONICYT (Chile), the Australian Research Council (Australia),
Minist\'{e}rio da Ci\^{e}ncia, Tecnologia e Inova\c{c}\~{a}o (Brazil) and
Ministerio de Ciencia, Tecnolog\'{i}a e Innovaci\'{o}n Productiva (Argentina).
2MASS is a joint project of the University of
Massachusetts and the Infrared Processing and Analysis Center (IPAC) at
Caltech, funded by NASA and the NSF.
The Center for Exoplanets and Habitable Worlds is supported by the
Pennsylvania State University, the Eberly College of Science, and the
Pennsylvania Space Grant Consortium.

\clearpage

\begin{deluxetable}{ll}
\tabletypesize{\scriptsize}
\tablewidth{0pt}
\tablecaption{Parallax, Proper Motion, and Photometry for
WISE~J085510.83$-$071442.5\label{tab:data}}
\tablehead{
\colhead{Parameter} & \colhead{Value} \\
}
\startdata
$\pi$ & $0.454\pm$0.045$\arcsec$ \\
$\mu_{\alpha}$ cos $\delta$ & $-8.06\pm0.09\arcsec$~yr$^{-1}$ \\
$\mu_{\delta}$ & $0.70\pm0.07\arcsec$~yr$^{-1}$ \\
$J$ & $>23$\tablenotemark{a} \\
$K_s$ & $>18.6$\tablenotemark{a,b} \\
$W1$ & $>16.4$ \\
$W2$ & 13.89$\pm$0.05 \\
$W3$ & $\geq11.25$ \\
$W4$ & $>9$ \\
$[3.6]$ & 17.44$\pm$0.05 \\
$[4.5]$ & 13.89$\pm$0.02 \\
\enddata
\tablecomments{
These data apply to the moving component of WISE~J085510.83$-$071442.5.
The northern and southern stationary components have $[3.6]=16.70\pm0.04$
and $16.38\pm0.04$ and $[4.5]=16.08\pm0.04$ and $16.05\pm0.04$, respectively.}
\tablenotetext{a}{S/N$<$3.}
\tablenotetext{b}{Based on data from the VISTA Hemisphere Survey.}
\end{deluxetable}

\begin{deluxetable}{lllll}
\tabletypesize{\scriptsize}
\tablewidth{0pt}
\tablecaption{Astrometry for WISE~J085510.83$-$071442.5\label{tab:astro}}
\tablehead{
\colhead{$\alpha$ (J2000)} & \colhead{$\delta$ (J2000)} & \colhead{$\sigma_{\alpha,\delta}$} & \colhead{MJD} &  \colhead{Source} \\
\colhead{($\arcdeg$)} & \colhead{($\arcdeg$)} & \colhead{($\arcsec$)} & \colhead{} & \colhead{}}
\startdata
133.795224 & $-$7.245138 & 0.40 & 55320.4 & {\it WISE} \\
133.794261 & $-$7.245124 & 0.40 & 55511.4 & {\it WISE} \\
133.788165 & $-$7.244508 & 0.04 & 56464.5 & {\it Spitzer} \\
133.787085 & $-$7.244451 & 0.04 & 56677.3 & {\it Spitzer} 
\enddata
\tablecomments{
These data apply to the moving component of WISE~J085510.83$-$071442.5.
The two stationary components have $\alpha=133.794680$ and 133.795099,
$\delta=-7.244528$ and $-$7.245711, and $\sigma_{\alpha,\delta}=0\farcs08$
in the {\it Spitzer} images.}
\end{deluxetable}

\clearpage

\end{document}